# Enhanced metal-insulator transition in freestanding VO$_2$ down to 5 nm thickness

Kun Han[1,2,†], Liang Wu[1,3,†,*], Yu Cao[4,†], Hanyu Wang[5], Chen Ye[1], Ke Huang[1], M. Motapothula[6,7], Hongna Xing[8], Xinghua Li[8], Dong-Chen Qi[9], Xiao Li[5], X. Renshaw Wang[1,10,*]

[1]*Division of Physics and Applied Physics, School of Physical and Mathematical Sciences, Nanyang Technological University, 21 Nanyang Link, 637371, Singapore*

[2]*Key Laboratory of Structure and Functional Regulation of Hybrid Materials of Ministry of Education, Institutes of Physical Science and Information Technology, Anhui University, Hefei 230601, China*

[3]*School of Material Science and Engineering, Kunming University of Science and Technology, Kunming, Yunnan 650093, China*

[4]*Department of Electrical and Computer Engineering, National University of Singapore, 4 Engineering Drive 3, Singapore 117583*

[5]*NNU-SULI Thermal Energy Research Center, Center for Quantum Transport and Thermal Energy Science (CQTES), School of Physics and Technology, Nanjing Normal University, Nanjing 210023, China*

[6]*Department of Physics and Astronomy, Uppsala University, Box 516, SE-75120 Uppsala, Sweden*

[7]*Department of Physics, SRM University AP, Amaravati, Andhra Pradesh 522-502, India*

[8]*School of Physics, Northwest University, Xi'an 710069, China*

[9]*Centre for Materials Science, School of Chemistry and Physics, Queensland University of Technology, Brisbane, QLD 4001, Australia*

[10]*School of Electrical and Electronic Engineering, Nanyang Technological University, 50 Nanyang Ave, 639798, Singapore*

[†]These authors contributed equally

*e-mail: liangwu@kust.edu.cn, renshaw@ntu.edu.sg

**ABSTRACT**

Ultrathin freestanding membranes with a pronounced metal-insulator transition (MIT) provides huge potential in future flexible electronic applications as well as a unique aspect of the study of lattice-electron interplay. However, the reduction of the thickness to an ultrathin region (a few nm) is typically detrimental to the MIT in epitaxial films, and even catastrophic for their freestanding form. Here, we report an enhanced MIT in $VO_2$-based freestanding membranes, with a lateral size up to millimetres and $VO_2$ thickness down to 5 nm. The $VO_2$-membranes were detached by dissolving a $Sr_3Al_2O_6$ sacrificial layer between the $VO_2$ thin film and $c$-$Al_2O_3$(0001) substrate, allowing a transfer onto arbitrary surfaces. Furthermore, the MIT in the $VO_2$-membrane was greatly enhanced by inserting an intermediate $Al_2O_3$ buffer layer. In comparison to the best available ultrathin $VO_2$-membranes, the enhancement of MIT is over 400% at 5 nm $VO_2$ thickness and more than one order of magnitude for $VO_2$ above 10 nm. Our study widens the spectrum of functionality in ultrathin and large-scale membranes, and enables the potential integration of MIT into flexible electronics and photonics.

**INTRODUCTION**

Metal-insulator transition (MIT), an appealing strongly correlated electronic phenomenon, offers switchable electronics states, which are ideal for future electronic and photonic applications[1]. $VO_2$ is considered as one of the most attractive MIT materials, owing to its near room transition temperature (~341 K), where electrical conductivity can change up to six orders of magnitude[2-3]. Most recently, spurred by the technological interest for flexible electronics, such as flexible artificial neuron[4-5] and low-power transistor[6]. It is of great demand to transform the MIT in $VO_2$ from a conventional epitaxial rigid form to a freestanding flexible one. Meaningful endeavours have achieved the $VO_2$-membranes (typically above 25 nm) using a number of techniques, including ion milling[7], wet-etching based method[8-10], or mechanical exfoliation from mica substrate[11]. Furthermore, thanks to the continuous advancement of fabrication techniques, reducing the dimensions while retaining high-quality MIT in a flexible form can potentially lead to new insights into the interplays of lattice and electronic correlation[12] and utilization of flexible form of MIT in low-power flexible electronics. Hence, it is highly desirable to adopt a facile and gentle approach to preserve the MIT in freestanding large-scale $VO_2$ down to an ultrathin region, *i.e.* below 10 nm[7, 13].

However, it is extremely challenging to obtain a freestanding ultrathin $VO_2$ membrane (< 10 nm) with high-quality MIT. This is due to a combination of several fundamental and technical issues, including the fundamental limit of critical thickness[14], crystalline defects created during the fabrication[13], and unwanted surface states introduced during the sample processing[8-10]. To date, only the complicated technique of ion milling[7] managed to achieve a sub-20 nm $VO_2$-membranes by compromising the lateral size to tens of micrometres (limited by the growth approach) and a degraded MIT (due to ion-induced structural damages).

In this study, we demonstrate a facile fabrication to retain enhanced MIT in ultrathin and large-scale $VO_2$-membranes by utilizing an $Al_2O_3$ buffer layer and a water-soluble sacrificial $Sr_3Al_2O_6$ (SAO) layer[12, 15-17], albeit the underneath cubic SAO experiences a non-epitaxial growth on hexagonal *c*-$Al_2O_3$(0001) substrate. More importantly, by inserting an intermediate $Al_2O_3$ buffer layer between the SAO and $VO_2$ films, the millimetre-scale $VO_2$/$Al_2O_3$-membranes exhibit an enhanced MIT with the $VO_2$ thickness down to 5 nm. Quantitatively, this enhanced MIT is over 400% at 5 nm $VO_2$ thickness and more than one order of magnitude for $VO_2$ above 10 nm in comparison to the existing ultrathin $VO_2$-membranes[7].

**Experimental section**

Sample fabrication

Single monoclinic (*M*)-phase $VO_2$ films with thickness from 4 to 120 nm were grown by pulsed laser deposition (PLD) on *c*-$Al_2O_3$ substrates with/without an SAO sacrificial layer and/or an $Al_2O_3$ buffer layer between them. During the deposition, the laser fluence was fixed at ~2 J/cm$^2$ with a repetition rate of 5 Hz. The growth temperature and oxygen partial pressure ($P_{O2}$) for SAO and $Al_2O_3$ were 700 °C and $1 \times 10^{-5}$ Torr, respectively. Then the temperature was decreased to 500 °C under the same $P_{O2}$. $VO_2$ thin film was deposited at 500 °C with a $P_{O2}$ of $2 \times 10^{-4}$ Torr. After deposition, the sample was *in situ* annealed at 500 °C with a $P_{O2}$ of $5 \times 10^{-3}$ Torr for 1 h. Sequentially, the sample was cooled down to room temperature under the same $P_{O2}$. Sintered polycrystalline SAO, $Al_2O_3$ ceramic pellets and a commercial vanadium single crystal were used as targets. The thickness was controlled by the number of the laser pulse, and further calibrated by both stylus profilometer and X-ray reflectivity (see **Figure S1** in supporting information) Two types of heterostructure are fabricated, namely VO$_2$/SAO/*c*-Al$_2$O$_3$ and VO$_2$/Al$_2$O$_3$/SAO/*c*-Al$_2$O$_3$ heterostructures. As the SAO layer is dissolvable in deionized (DI) water[15, 18], immersing the heterostructures can detach the VO$_2$- or VO$_2$/Al$_2$O$_3$-membranes from the *c*-Al$_2$O$_3$ substrates, and get suspended in the DI water, which are ready to be transferred onto any arbitrary substrates. At last, in order to ensure a firm adhesion of the membranes onto the new surface of substrates, *e.g.* glass, the transferred membranes with the new substrates were annealed in a vacuum (< 10$^{-6}$ Torr) at 120 °C for 1 h to remove the bubbles and residual water.

Characterization methods

The electrical transport properties were measured with a Quantum Design physical property measurement system (PPMS) in the temperature range from 300 to 400 K. X-ray diffraction (XRD) was performed on a Bruker D8 diffractometer equipped with a Cu K$_{\alpha 1}$ source at a wavelength of 1.5406 Å operated at 40 keV and 40 mA in grazing incidence (GI) geometry. Raman spectra were performed in backscattering configuration and recorded at 532 nm laser excitation. The resolution for this configuration is 0.5 cm$^{-1}$.

**RESULTS AND DISCUSSION**

**Figure 1** shows the deposition, detachment and transfer processes to achieve the VO$_2$ and VO$_2$/Al$_2$O$_3$ membranes. The MIT in VO$_2$ is accompanied by a structural phase transition from a low-temperature *M*-phase to a high-temperature rutile (*R*)-phase. **Figure 1** also shows optical images of the VO$_2$- and VO$_2$/Al$_2$O$_3$-membranes on glass substrates with a scale up to several millimetres.

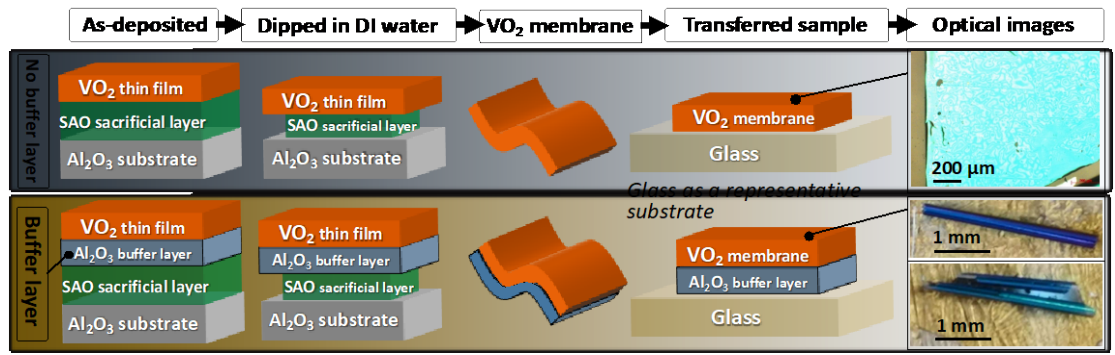

**Figure 1. Schematics of the process to obtain ultrathin and large-scale VO$_2$ and VO$_2$/Al$_2$O$_3$ membranes.** The as-grown samples, namely VO$_2$/SAO/$c$-Al$_2$O$_3$ and VO$_2$/Al$_2$O$_3$/SAO/$c$-Al$_2$O$_3$ heterostructures, are dipped in DI water to remove the SAO sacrificial layer. Consequently, ultrathin VO$_2$- and VO$_2$/Al$_2$O$_3$-membranes are achieved and can be transferred onto any substrates. In the schematics, glass substrates were used as a representative substrate. Eventually, optical images of the ultrathin and large-scale VO$_2$- and VO$_2$/Al$_2$O$_3$-membranes with a thickness down to 5 nm and a feature-length of mm are captured. The difference between these two types of membranes is the insertion of a buffer layer of Al$_2$O$_3$ during the growth.

It is worth mentioning that the efficiency of dissolving the SAO sacrificial layer is dramatically increased after inserting the Al$_2$O$_3$ buffer layer. The durations required for dissolving of the SAO layers in the 5 mm × 5 mm of VO$_2$/SAO/$c$-Al$_2$O$_3$ and VO$_2$/Al$_2$O$_3$/SAO/$c$-Al$_2$O$_3$ heterostructures are 1440 and 10 min, respectively. Moreover, the VO$_2$/Al$_2$O$_3$ membrane can be rolled into microtube during the dissolution of the SAO sacrificial layer in DI water without breaking, suggesting that the VO$_2$-membrane is highly flexible to bear a large degree of deformation. Specifically, the VO$_2$/Al$_2$O$_3$-membrane remains flat when the thickness of Al$_2$O$_3$ is below 10 nm and rolls up into a microtube for thicker Al$_2$O$_3$. The self-bending is ascribed to the low bending stiffness of the bilayer membrane system[16] and unreleased strain between the Al$_2$O$_3$ and VO$_2$ layer.

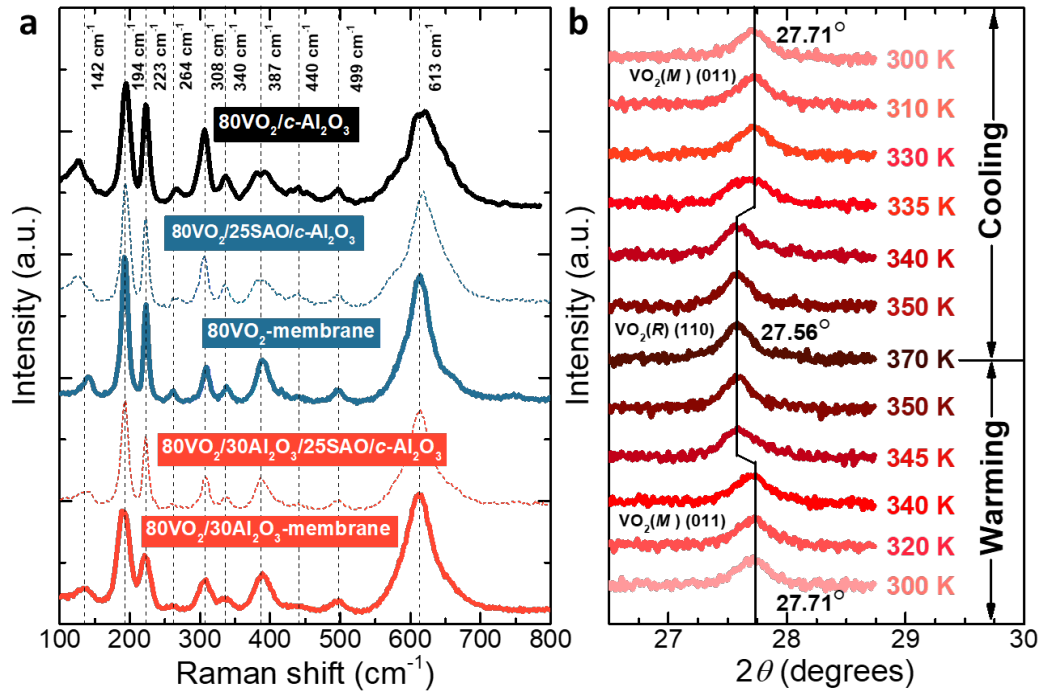

**Figure 2. Structural characterisation of VO$_2$-based heterostructures and membranes.** (**a**) Room temperature Raman spectra of 80VO$_2$/c-Al$_2$O$_3$, 80VO$_2$/25SAO/c-Al$_2$O$_3$, 80VO$_2$/30Al$_2$O$_3$/25SAO/c-Al$_2$O$_3$, 80VO$_2$-membrane, 80VO$_2$/30Al$_2$O$_3$-membrane. (**b**) Temperature-dependent HRXRD 2θ-ω scans of VO$_2$ (011) peak of the 80VO$_2$/30Al$_2$O$_3$/25SAO/c-Al$_2$O$_3$ at temperatures ranging from 300 to 370 K. The black arrow lines denote the peak shift across the MIT.

The structural phase of all the epitaxial films and membranes were firstly characterized by Raman spectroscopy with a 532 nm excitation laser source and temperature-dependent high-resolution X-ray diffraction (HRXRD). **Figure 2a** shows the room temperature Raman scattering curves for the 80 nm VO$_2$/c-Al$_2$O$_3$ heterostructure (abbreviated as 80VO$_2$/c-Al$_2$O$_3$, same notion will be used thereafter), 80VO$_2$/25SAO/c-Al$_2$O$_3$, 80VO$_2$-membrane, 80VO$_2$/30Al$_2$O$_3$/25SAO/c-Al$_2$O$_3$, and 80VO$_2$/30Al$_2$O$_3$-membrane. All the peaks at 194, 223, 264, 308, 340, 387, 440, 499, and 613 cm$^{-1}$ correspond to the pure *M*-phase VO$_2$ in both heterostructure and membrane forms at room temperature[19]. Moreover, the HRXRD measurement was conducted on VO$_2$/Al$_2$O$_3$/SAO/c-Al$_2$O$_3$ to confirm the thermal driven first-order SPT. **Figure 2b** shows that the VO$_2$(*M*) (011) peaks at 27.72° in the low-temperature region, *i.e.* from 300 to 340 K, shifts to rutile VO$_2$(*R*) (110) peaks at 27.58° in the high-temperature region, *i.e.* from 345 to 370 K. When subsequently cooling the samples, the VO$_2$(R) (110) peaks recover to the initial VO$_2$(M) (011) peaks position[20-21]. Therefore, both Raman and HRXRD results demonstrated that the crystallinity of the VO$_2$(M) are well-retained when grown on SAO, and which is barely affected by the transfer process.

The electrical properties of the VO$_2$ heterostructures and membranes were characterized by the temperature-dependent resistivity and corresponding derivative curves, d(log$\rho$)/dT. **Figure 3** shows that the MIT behaviour was preserved in all VO$_2$/SAO/c-Al$_2$O$_3$ samples, though the steepness of resistivity change is compromised and the magnitude of the resistivity change is suppressed in contrast to the epitaxial VO$_2$/c-Al$_2$O$_3$. The degraded MIT behaviour of VO$_2$/SAO/c-Al$_2$O$_3$ can be mainly attributed to the defects in VO$_2$ thin film induced by the SAO sacrificial layer. To overcome this, an Al$_2$O$_3$ buffer layer was inserted between VO$_2$ thin film and SAO sacrificial layer. **Figure 3a** shows that the resistivity change of VO$_2$/Al$_2$O$_3$/SAO/c-Al$_2$O$_3$ increases by around one order of magnitude and the MIT becomes much steeper. **Figure 3b** shows the normalized resistivity change, *i.e.* $\rho(T)/\rho(300\ K)$, as a function of temperature. Notably, the magnitude of the resistivity changes for VO$_2$/SAO/c-Al$_2$O$_3$ and VO$_2$-membrane are comparable, as well as for VO$_2$/Al$_2$O$_3$/SAO/c-Al$_2$O$_3$ and VO$_2$/Al$_2$O$_3$-membrane. The detailed comparison of MIT of different VO$_2$ can be found in **Table S1** in supporting information. The electrical transport measurements suggest that the influence of transferring procedure on its electrical transport properties is negligible.

To further characterize the change of MIT transition temperature ($T_{MIT}$) and the sharpness of resistivity change, several parameters are defined and plotted in **Figure 3c**. $T_{Heat}$ and $T_{Cool}$ are the specific transition temperature defined as the corresponding peak position of the derivative curves, d(log$\rho$)/dT, during heating and cooling, respectively. Hence, $T_{MIT}$ can be defined as $T_{MIT} = (T_{Heat} + T_{Cool})/2$. The hysteresis width ($\Delta H$) is defined as the difference between $T_{Heat}$ and $T_{Cool}$. The sharpness of the transition is defined as the full width at half maximum (FWHM) of the derivative curves during heating and cooling, which is derived by Gaussian fitting. **Figure 3c** shows that both $T_{Heat}$ and $T_{Cool}$ are slightly shifted towards room temperature for VO$_2$/SAO/c-Al$_2$O$_3$ and VO$_2$/Al$_2$O$_3$/SAO/c-Al$_2$O$_3$ and their corresponding membranes by comparing with that of VO$_2$/c-Al$_2$O$_3$. The decrease of $T_{MIT}$ indicates VO$_2$ is compressively strained in both heterostructure and membranes forms by inserting the SAO sacrificial layer and

Al$_2$O$_3$ buffer layer[9, 22]. The FWHM becomes much broader for VO$_2$/SAO/*c*-Al$_2$O$_3$ and VO$_2$/SAO-membrane, suggesting the degraded quality of VO$_2$[10]. After inserting an Al$_2$O$_3$ buffer layer between VO$_2$ and SAO, the resistivity change becomes sharper[23], which again proves the beneficial effect of buffer Al$_2$O$_3$.

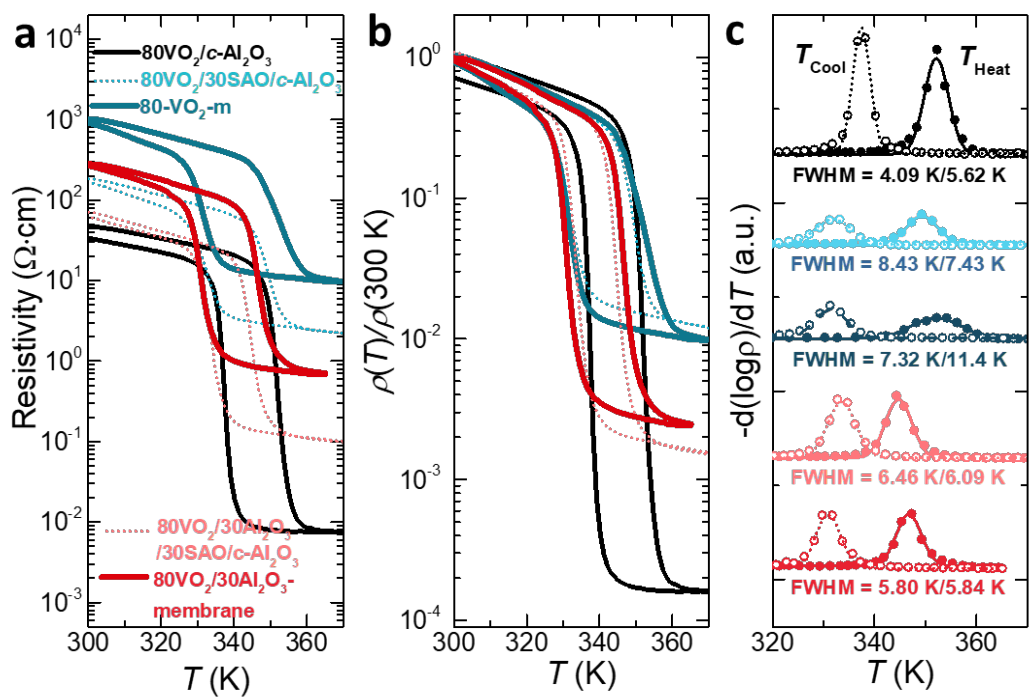

**Figure 3. Effect of Al$_2$O$_3$ buffer layer.** (**a**) Resistivity, (**b**) normalized resistivity changes, *i.e.* $\rho(T)/\rho(300\,K)$, and (**c**) derivatives of $\log_{10}\rho(T)$ as a function of temperature during the cooling (hollow circle) and heating (solid circle) cycles of the various VO$_2$-based heterostructures and membranes. The symbols and lines represent data points and Gaussian fitting curves, respectively.

The MIT properties of VO$_2$ are dramatically affected by thickness as well[7, 24]. A minimum thickness is needed for the full development of the electronic MIT, such as the study of quantum confinement effects[7]. Therefore, an ultrathin VO$_2$ membrane is a prerequisite to have a deeper examination of the intrinsic MIT properties and quantum size effects. **Figure 4a-e** shows temperature-dependent resistivity of VO$_2$/*c*-Al$_2$O$_3$, VO$_2$/SAO/*c*-Al$_2$O$_3$, VO$_2$-membrane, VO$_2$/Al$_2$O$_3$/SAO/*c*-Al$_2$O$_3$, and VO$_2$/Al$_2$O$_3$-membrane with different VO$_2$ thicknesses. For all VO$_2$ films with thickness above 20 nm, the MIT properties are retained. It is worth noting that the critical thickness for MIT in VO$_2$ film increases from 4 nm for VO$_2$/*c*-Al$_2$O$_3$ to 15 nm for VO$_2$/SAO/*c*-Al$_2$O$_3$ and 20 nm for VO$_2$-membrane, indicating that the VO$_2$ deteriorates when grown on SAO. Remarkably, the critical thickness recovers to 5 nm for VO$_2$/Al$_2$O$_3$/SAO/*c*-Al$_2$O$_3$ and corresponding VO$_2$/Al$_2$O$_3$-membrane, demonstrating that the quality of the VO$_2$ is comparable with that of VO$_2$ directly grown on *c*-Al$_2$O$_3$ substrate[25].

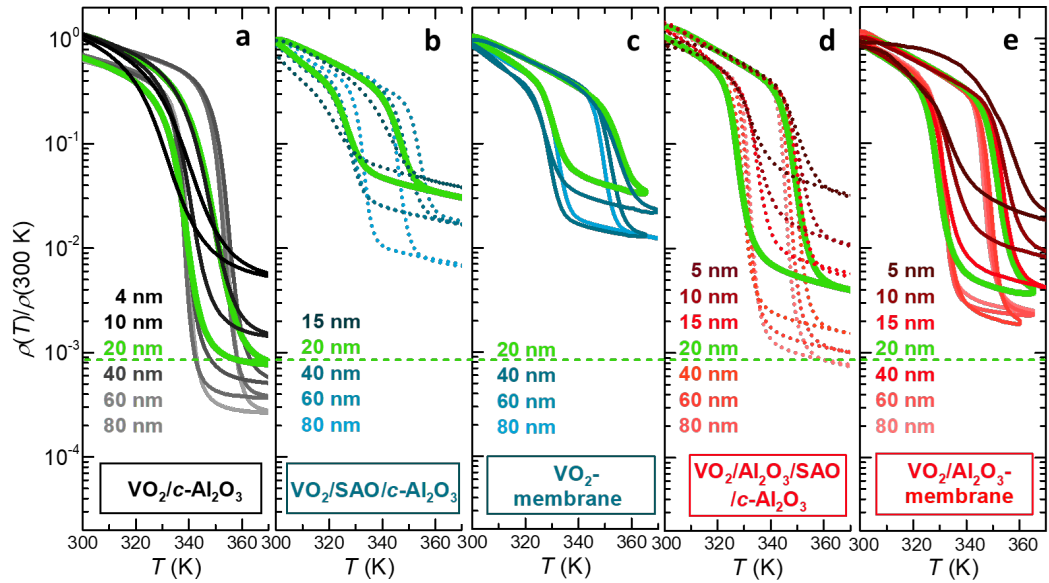

**Figure 4. Thickness dependent metal-insulator transition of VO$_2$-based heterostructures and VO$_2$-membranes.** Normalized resistivity change, *i.e.* $\rho(T)/\rho(300\ K)$, of (**a**) VO$_2$/*c*-Al$_2$O$_3$, (**b**) VO$_2$/SAO/*c*-Al$_2$O$_3$, (**c**) VO$_2$-membrane, (**d**) VO$_2$/Al$_2$O$_3$/SAO/*c*-Al$_2$O$_3$, and (**e**) VO$_2$/Al$_2$O$_3$-membrane. The labels of the thickness in all figures correspond to the VO$_2$ thicknesses in the respective samples.

**Figure 5** compares the thickness dependence of the resistivity change across the MIT, $\rho(300\ K)/\rho(370\ K)$ of different ultrathin VO$_2$-membranes prepared in this work and the pioneer work in ref [7]. One can clearly see the magnitude of the resistivity change during the MIT in this study is over one order of magnitude better than the previous study for VO$_2$ above 10 nm, which can still be maintained to 400% when the thickness of VO$_2$ is reduced to 5 nm. In addition to the exceptional MIT, the water-soluable sacrificial SAO layer approach shows celebrated advantages in obtaining both large-scale, high-quality VO$_2$ membranes with an easy-handling process.

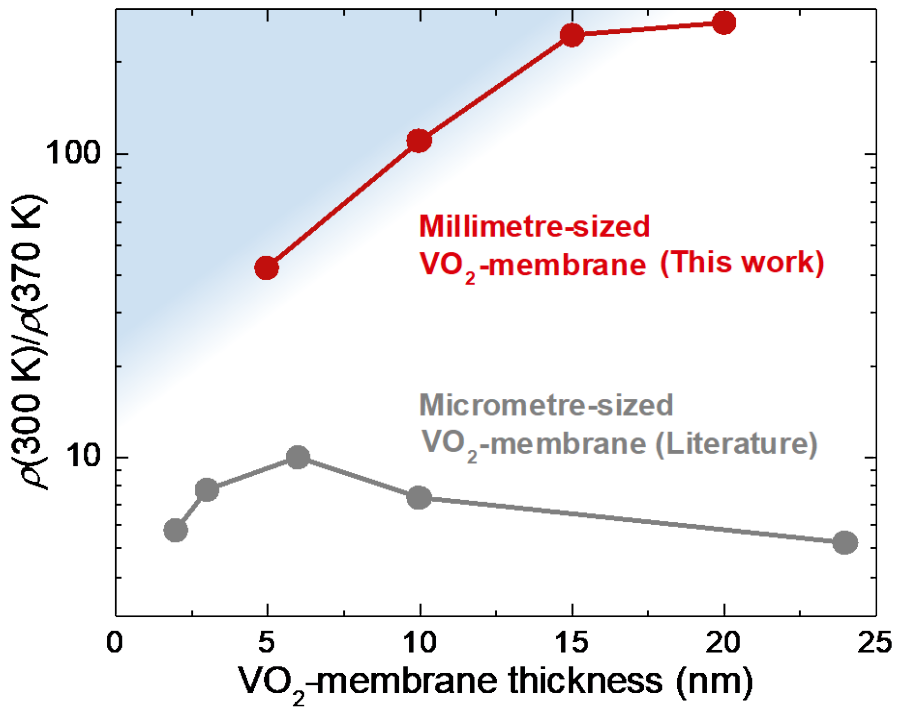

**Figure 5. Comparison of the MIT in ultrathin VO$_2$-membranes (less than 25 nm) fabricated by different methods.** The thickness dependence of the resistivity change, $\rho$(300 K)/$\rho$(370 K), across the MIT of VO$_2$-membranes by different fabrication methods.

## Conclusion

In summary, we have demonstrated a facile synthesis method to fabricate freestanding large-scale VO$_2$ ultrathin membranes by inserting SAO as a sacrificial layer and Al$_2$O$_3$ as a buffer layer between the VO$_2$ film and *c*-Al$_2$O$_3$ substrate. Both the *M*-phase crystal structure and the corresponding MIT behaviour are preserved in the VO$_2$- and VO$_2$/Al$_2$O$_3$-membranes form, albeit the SAO cannot be epitaxially grown on *c*-Al$_2$O$_3$ substrate. Furthermore, the MIT quality of the VO$_2$ membranes is notably improved by employing an Al$_2$O$_3$ buffer layer even for VO$_2$ down to 5 nm, which shows great advantages as compared with the previous studies and guarantees the excellent flexibility for future low-power flexible electronics. Our VO$_2$/Al$_2$O$_3$-membranes demonstrated advantages of the higher magnitude of the MIT resistivity change in a larger area and thinner VO$_2$-membranes than the VO$_2$-membranes prepared by other techniques. Our results could pave the way for practically and versatilely integrating high-quality MIT in future flexible electronics and photonics.

**Supporting Information**

Detailed comparison of MIT of different $VO_2$ films (Table S1); Atomic force microscopy and X-ray reflectivity calibration of film thickness (Figure S1)

**Acknowledgement**

X.R.W. acknowledges supports from the Academic Research Fund Tier 1 (Grant No. RG177/18) from Singapore Ministry of Education, and the Singapore National Research Foundation (NRF) under the competitive Research Programs (CRP Grant No. NRF-CRP21-2018-0003), and Agency for Science, Technology and Research (A*STAR) under its AME IRG grant (Project No. A20E5c0094). D.Q. acknowledges the support of the Australian Research Council (Grant No. FT160100207) and the continued support from the Queensland University of Technology (QUT) through the Centre for Materials Science. K. H. acknowledges the support from Singapore National Research Foundation (NRF) under the Competitive Research Programs (CRP Award No. NRFCRP15-2015-01). We thank C. Yu, SW Zeng, Z. Huang, S. Goswami, A. Ariando and T. Venky Venkatasen on their fruitful discussion and help during the sample growth.

**REFERENCES**

(1) Imada, M.; Fujimori, A.; Tokura, Y. Metal-Insulator Transitions. *Rev. Mod. Phys.* **1998,** *70* (4), 1039-1263.
(2) Shao, Z.; Cao, X.; Luo, H.; Jin, P. Recent Progress in the Phase-Transition Mechanism and Modulation of Vanadium Dioxide Materials. *NPG Asia Mater.* **2018,** *10* (7), 581-605.
(3) Liu, K.; Lee, S.; Yang, S.; Delaire, O.; Wu, J. Recent Progresses on Physics and Applications of Vanadium Dioxide. *Mater. Today* **2018,** *21* (8), 875-896.
(4) Yi, W.; Tsang, K. K.; Lam, S. K.; Bai, X.; Crowell, J. A.; Flores, E. A. Biological Plausibility and Stochasticity in Scalable $VO_2$ Active Memristor Neurons. *Nat. Commun.* **2018,** *9* (1), 4661.
(5) del Valle, J.; Salev, P.; Tesler, F.; Vargas, N. M.; Kalcheim, Y.; Wang, P.; Trastoy, J.; Lee, M.-H.; Kassabian, G.; Ramírez, J. G.; Rozenberg, M. J.; Schuller, I. K. Subthreshold Firing in Mott Nanodevices. *Nature* **2019,** *569* (7756), 388-392.
(6) Shukla, N.; Thathachary, A. V.; Agrawal, A.; Paik, H.; Aziz, A.; Schlom, D. G.; Gupta, S. K.; Engel-Herbert, R.; Datta, S. A Steep-Slope Transistor Based on Abrupt Electronic Phase Transition. *Nat. Commun.* **2015,** *6* (1), 7812.
(7) Fadlelmula, M. M.; Sürmeli, E. C.; Ramezani, M.; Kasırga, T. S. Effects of Thickness on the Metal–Insulator Transition in Free-Standing Vanadium Dioxide Nanocrystals. *Nano Lett.* **2017,**

*17* (3), 1762-1767.

(8) Tian, Z.; Xu, B.; Hsu, B.; Stan, L.; Yang, Z.; Mei, Y. Reconfigurable Vanadium Dioxide Nanomembranes and Microtubes with Controllable Phase Transition Temperatures. *Nano Lett.* **2018,** *18* (5), 3017-3023.

(9) Sim, J. S.; Zhou, Y.; Ramanathan, S. Suspended Sub-50 nm Vanadium Dioxide Membrane Transistors: Fabrication and Ionic Liquid Gating Studies. *Nanoscale* **2012,** *4* (22), 7056-62.

(10) Pellegrino, L.; Manca, N.; Kanki, T.; Tanaka, H.; Biasotti, M.; Bellingeri, E.; Siri, A. S.; Marre, D. Multistate Memory Devices Based on Free-Standing $VO_2/TiO_2$ Microstructures Driven by Joule Self-Heating. *Adv. Mater.* **2012,** *24* (21), 2929-34.

(11) Chen, Y.; Fan, L.; Fang, Q.; Xu, W.; Chen, S.; Zan, G.; Ren, H.; Song, L.; Zou, C. Free-Standing $SWNTs/VO_2$/Mica Hierarchical Films for High-Performance Thermochromic Devices. *Nano Energy* **2017,** *31*, 144-151.

(12) Hong, S. S.; Gu, M.; Verma, M.; Harbola, V.; Wang, B. Y.; Lu, D.; Vailionis, A.; Hikita, Y.; Pentcheva, R.; Rondinelli, J. M.; Hwang, H. Y. Extreme Tensile Strain States in $La_{0.7}Ca_{0.3}MnO_3$ Membranes. *Science* **2020,** *368* (6486), 71-76.

(13) Peter, A. P.; Martens, K.; Rampelberg, G.; Toeller, M.; Ablett, J. M.; Meersschaut, J.; Cuypers, D.; Franquet, A.; Detavernier, C.; Rueff, J.-P.; Schaekers, M.; Van Elshocht, S.; Jurczak, M.; Adelmann, C.; Radu, I. P. Metal-Insulator Transition in ALD $VO_2$ ultrathin Films and Nanoparticles: Morphological Control. *Adv. Funct. Mater.* **2015,** *25* (5), 679-686.

(14) Hong, S. S.; Yu, J. H.; Lu, D.; Marshall, A. F.; Hikita, Y.; Cui, Y.; Hwang, H. Y. Two-Dimensional Limit of Crystalline Order in Perovskite Membrane Films. *Sci. Adv.* **2017,** *3* (11), eaao5173.

(15) Lu, D.; Baek, D. J.; Hong, S. S.; Kourkoutis, L. F.; Hikita, Y.; Hwang, Harold Y. Synthesis of Freestanding Single-Crystal Perovskite Films and Heterostructures by Etching of Sacrificial Water-Soluble layers. *Nat. Mater.* **2016,** *15* (12), 1255-1260.

(16) Dong, G.; Li, S.; Yao, M.; Zhou, Z.; Zhang, Y. Q.; Han, X.; Luo, Z.; Yao, J.; Peng, B.; Hu, Z.; Huang, H.; Jia, T.; Li, J.; Ren, W.; Ye, Z. G.; Ding, X.; Sun, J.; Nan, C. W.; Chen, L. Q.; Li, J.; Liu, M. Super-Elastic Ferroelectric Single-Crystal Membrane with Continuous Electric Dipole Rotation. *Science* **2019,** *366* (6464), 475-479.

(17) Ji, D.; Cai, S.; Paudel, T. R.; Sun, H.; Zhang, C.; Han, L.; Wei, Y.; Zang, Y.; Gu, M.; Zhang, Y.; Gao, W.; Huyan, H.; Guo, W.; Wu, D.; Gu, Z.; Tsymbal, E. Y.; Wang, P.; Nie, Y.; Pan, X. Freestanding Crystalline Oxide Perovskites Down to the Monolayer Limit. *Nature* **2019,** *570* (7759), 87-90.

(18) Han, K.; Hu, K.; Li, X.; Huang, K.; Huang, Z.; Zeng, S.; Qi, D.; Ye, C.; Yang, J.; Xu, H.; Ariando, A.; Yi, J.; Lu, W.; Yan, S.; Wang, X. R. Erasable and Recreatable Two-Dimensional Electron Gas at the Heterointerface of $SrTiO_3$ and a Water-Dissolvable Overlayer. *Sci. Adv.* **2019,** *5* (8), eaaw7286.

(19) Chen, F. H.; Fan, L. L.; Chen, S.; Liao, G. M.; Chen, Y. L.; Wu, P.; Song, L.; Zou, C. W.; Wu, Z. Y. Control of the Metal-Insulator Transition in $VO_2$ Epitaxial Film by Modifying Carrier Density. *ACS Appl. Mater. Interfaces* **2015,** *7* (12), 6875-81.

(20) Hada, M.; Okimura, K.; Matsuo, J. Characterization of Structural Dynamics of $VO_2$ Thin Film on *c*-$Al_2O_3$ Using in-Air Time-Resolved X-Ray Diffraction. *Phys. Rev. B* **2010,** *82* (15), 153401.

(21) Yang, M.; Yang, Y.; Wang, L.; Hong, B.; Huang, H.; Hu, X.; Zhao, Y.; Dong, Y.; Li, X.; Lu,

Y.; Bao, J.; Luo, Z.; Gao, C. For Progress in Natural Science: Materials International Investigations of Structural Phase Transformation and THz Properties across Metal–Insulator Transition in $VO_2$/ $Al_2O_3$ Epitaxial Films. *Prog. Nat. Sci-Mater* **2015,** *25* (5), 386-391.

(22) Muraoka, Y.; Hiroi, Z. Metal–Insulator Transition of $VO_2$ Thin Films Grown on $TiO_2$ (001) and (110) Substrates. *Appl. Phys. Lett.* **2002,** *80* (4), 583-585.

(23) Lee, D.; Lee, J.; Song, K.; Xue, F.; Choi, S. Y.; Ma, Y.; Podkaminer, J.; Liu, D.; Liu, S. C.; Chung, B.; Fan, W.; Cho, S. J.; Zhou, W.; Lee, J.; Chen, L. Q.; Oh, S. H.; Ma, Z.; Eom, C. B. Sharpened $VO_2$ Phase Transition Via Controlled Release of Epitaxial Strain. *Nano Lett* **2017,** *17* (9), 5614-5619.

(24) Yamin, T.; Wissberg, S.; Cohen, H.; Cohen-Taguri, G.; Sharoni, A. Ultrathin Films of $VO_2$ on r-Cut Sapphire Achieved by Postdeposition Etching. *ACS Appl. Mater. Interfaces* **2016,** *8* (23), 14863-70.

(25) Zhang, H.-T.; Zhang, L.; Mukherjee, D.; Zheng, Y.-X.; Haislmaier, R. C.; Alem, N.; Engel-Herbert, R. Wafer-Scale Growth of $VO_2$ Thin Films Using a Combinatorial Approach. *Nat. Commun.* **2015,** *6* (1), 8475.

**Supporting Information**

**Enhanced metal-insulator transition in freestanding VO$_2$ down to 5 nm thickness**

Kun Han[1,2, †], Liang Wu[1,3, † *], Yu Cao[4, †], Hanyu Wang[5], Chen Ye[1], Ke Huang[1], M. Motapothula[6,7], Hongna Xing[8], Xinghua Li[8], Dong-Chen Qi[9], Xiao Li[5], X. Renshaw Wang[1,10,*]

[1]*Division of Physics and Applied Physics, School of Physical and Mathematical Sciences, Nanyang Technological University, 21 Nanyang Link,637371, Singapore*

[2]*Key Laboratory of Structure and Functional Regulation of Hybrid Materials of Ministry of Education, Institutes of Physical Science and Information Technology, Anhui University, Hefei 230601, China*

[3]*School of Material Science and Engineering, Kunming University of Science and Technology, Kunming, Yunnan 650093, China*

[4]*Department of Electrical and Computer Engineering, National University of Singapore, 4 Engineering Drive 3, Singapore 117583*

[5]*NNU-SULI Thermal Energy Research Center, Center for Quantum Transport and Thermal Energy Science (CQTES), School of Physics and Technology, Nanjing Normal University, Nanjing 210023, China*

[6]*Department of Physics and Astronomy, Uppsala University, Box 516, SE-75120 Uppsala, Sweden*

[7]*Department of Physics, SRM University AP, Amaravati, Andhra Pradesh 522-502, India*

[8]*School of Physics, Northwest University, Xi'an 710069, China*

[9]*Centre for Materials Science, School of Chemistry and Physics, Queensland University of Technology, Brisbane, QLD 4001, Australia*

[10]*School of Electrical and Electronic Engineering, Nanyang Technological University, 50 Nanyang Ave, 639798, Singapore*

†These authors contributed equally

*e-mail: liangwu@kust.edu.cn, renshaw@ntu.edu.sg

## Table S1

To quantitatively evaluate the quality of our VO₂ films and membranes, several key parameters are defined and listed in **Table S1**.

| | $\Delta A$ | $T_{heat}$ (K) | $T_{cool}$ (K) | $T_{MIT}$ (K) | $\Delta H$ (K) | $\Delta T$ (K) |
|---|---|---|---|---|---|---|
| VO₂/c-Al₂O₃ | 6223 | 352.3 | 337.2 | 344.8 | 15.1 | 4.9 |
| VO₂/SAO/c-Al₂O₃ | 138 | 345.9 | 333.2 | 339.6 | 12.7 | 7.4 |
| VO₂-membrane | 83 | 349.2 | 332.2 | 340.7 | 17 | 7.2 |
| VO₂/Al₂O₃/SAO/c-Al₂O₃ | 1238 | 345.8 | 332.5 | 339.2 | 13.3 | 5.2 |
| VO₂/Al₂O₃-membrane | 409 | 347 | 331 | 339 | 16 | 5.4 |

$\Delta A$: ($R(300K)/R(370K)$);
$T_{MIT} = (T_{heat} + T_{cool})/2$;
$T_{heat}$ and $T_{cool}$: the transition temperatures defined as the peak position of the derivative curves (d(log$\rho$)/d$T$) during heating and cooling.
$\Delta H$: The hysteresis width is defined $\Delta H = T_{heat} - T_{cool}$;
$\Delta T$: The sharpness of the transition, which is defined as the FWHM of the derivative curve during heating.

Figure S1

The thickness of VO2 films is calibrated by results of profilometry (**Figure S1a,b**) and X-ray reflectivity (**Figure S1c**). Both results are consistent, which provides the thickness controlling by counting the number of the laser pulse. For VO2, 1 nm corresponds to 500 pulses; For Al2O3, 1 nm corresponds to 20 pulses.

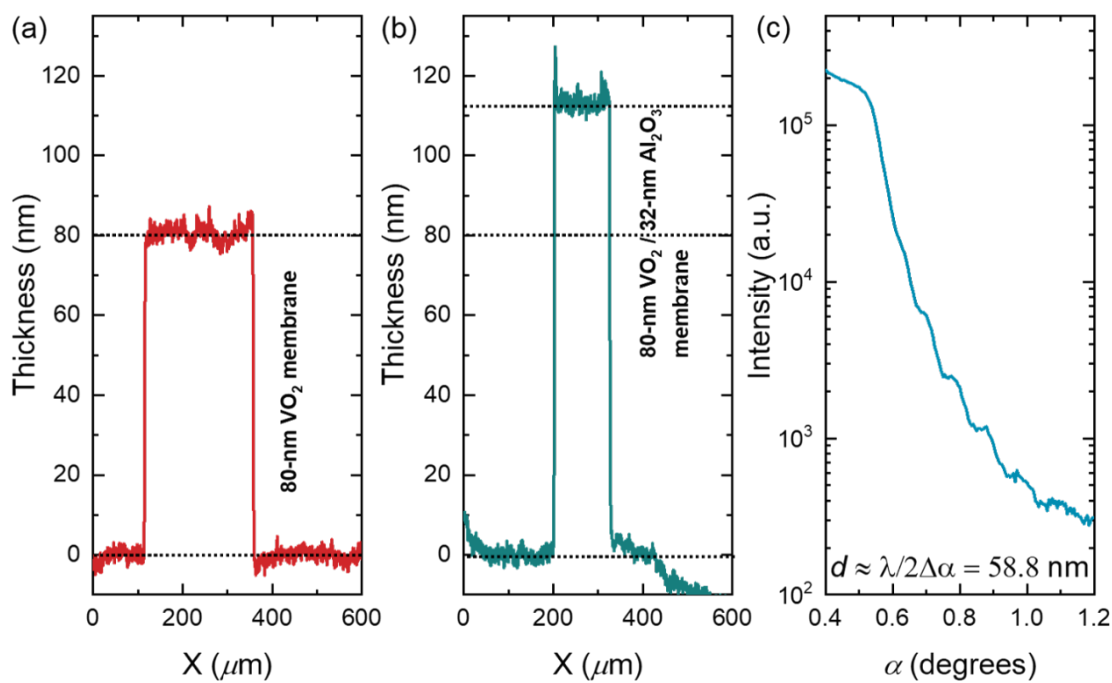

**Supplementary Figure S1. Thickness calibration of different VO2.** The thicknesses of (**a**) 40000 pulses VO2-membrane and (**b**) 40000 pulses VO2/600 pulses Al2O3 were measured by Bruker profilometer. (**c**) The X-ray reflectivity (XRR) of 30000 pulses VO2/c-Al2O3 heterostructure. The thickness of the VO2 film obtained by fitting is 58.8 nm. The thicknesses measured by both methods are consistent. All the thicknesses used in this study are calibrated by the laser pulses.